\documentclass[12pt]{iopart}
\usepackage{graphicx}
\usepackage{amsfonts}

\usepackage{tikz}
\usepackage{pgfplots}
\usetikzlibrary{matrix}

\begin{document}

\newlength\figureheight 
\newlength\figurewidth 

\bibliographystyle{unsrt}

\article[Computation of NLS minimal action paths]{}{Computation of minimum action paths of the stochastic nonlinear Schr{\"o}dinger equation with
dissipation}
\author{G. Poppe$^{1}$ and T. Sch{\"a}fer$^{1,2}$}
\address{$^1$Physics Program at the CUNY Graduate Center \\ $^2$Department of Mathematics, College of Staten Island}

\ead{tobias@math.csi.cuny.edu}

\begin{abstract}
  Using the geometric minimum action method, we compute minimizers
  of the Freidlin-Wentzell functional for the dissipative linear and nonlinear 
  Schr{\"o}dinger equation. For the particular case of transitions between
  solitary waves of different amplitudes, we discuss the relationship of 
  the minimizer of the PDE model to the minimizer of a finite-dimensional
  reduction.
\end{abstract}  
\maketitle

\noindent 
\section{Introduction}

Over the last decades, much effort has been made to develop analytical and computational tools to improve the understanding
of rare events in complex stochastic systems. Consider, for example, a classical dynamical system with two stable
fixed points where, in the absence of noise, the system is unable to switch from one fixed point to the other. In the presence
of noise, a transition from one stable fixed point to the other stable fixed point can become possible, but its occurrence will
be rare if the noise is small. When the switching occurs, the transition path is itself random and different transition paths have different likelihoods. 
The most probable path is the most important path, often called the {\em instanton}. In many cases, the probability distribution around the instanton
follows roughly $\sim {\mathrm{e}}^{-S/\epsilon}$ where the small parameter $\epsilon$ characterizes the strength of the noise and $S$ 
is the Freidlin-Wentzell functional \cite{freidlin-wentzell:1998} associated with the stochastic system. If the Freidlin-Wentzell theory is applicable,
the instanton can be found as the minimizer of $S$.

Given the fundamental importance of instantons for the noise-driven transition in stochastic systems, it is not surprising 
that they have been computed, described, and analyzed in a variety of contexts and fields - starting from the beginnings 
in the 70s (Martin-Siggia-Rose \cite{martin-siggia-rose:1973}, DeDominicis \cite{dominicis:1976}, Janssen \cite{janssen:1976}), 
to applications nowadays in many fields, including for example turbulence \cite{grafke-grauer-schaefer-etal:2015} and nonlinear optics
\cite{falkovich-etal:2001,terekhov-etal:2014}.
 While a great deal of work has been accomplished on the analytical side, the development of efficient numerical methods to compute 
 instantons, in particular in high-dimensional spaces, is more recent. Key steps on the computational side were the development
 of the {\em string method} \cite{e-ren-vanden-eijnden:2007} and the more general {\em geometric minimum action method} (gMAM) \cite{heymann-vanden-eijnden:2008,vanden-eijnden-heymann:2008}. Both methods can be used to compute instantons in instances where 
 the initial and the final state of the system are known. The string method is extremely successful in the case of diffusive processes where the drift field is a gradient field. The gMAM allows to handle non-gradient drift fields as well as more general noise processes. Note that the framework of gMAM is immediately applicable to stochastic ordinary differential equations, but it needs - like most numerical frameworks - to be adapted to the particular form of the stochastic partial differential equation under consideration.
 
 The present paper has two main objectives: First, it shows how to adapt and implement gMAM for a 
 stochastically driven nonlinear Schr{\"o}dinger equation. While, in terms of applications, we focus on the case of the transition of
 two solitary waves in the context of nonlinear optics, the presented gMAM for the nonlinear Schr{\"o}dinger equation can 
 find applications in a variety of areas, e.g. Bose-Einstein condensation or nonlinear wave phenomena. The second objective of
 the paper is to discuss the relationship of the instanton found in the stochastic PDE system to the instanton of an approximation given by a low-dimensional
stochastic dynamical system. This is important as, in the past, often such low-dimensional reductions have been used to analyze the behavior of 
 complex systems without being able to characterize the limitations of such an approach.  With the new tools developed in this paper it is now
 possible to look deeper into the validity of low-dimensional models.

\section{Review of the geometric minimum action method (gMAM)}

Consider a stochastic differential equation \cite{arnold:1974,oksendal:2003,gardiner:2009} for a $n$-dimensional stochastic process $x$ given by
\begin{equation} \label{basic_sde}
dx = b(x)dt + \sqrt{\epsilon}\sigma(x)dW\,.
\end{equation}
It is well-known that the transition probability (meaning the probability to find a particle at the location $\tilde x$ at the time $t$ 
assuming that the particle started at $x(0)=x_0$) can be written as a path integral \cite{langouche-roekaerts-etal:1982,kleinert:2009,Schafer_Moore_2011} in the form of 
\begin{equation}
p(\tilde x,t) = \int_{{\mathcal{C}}(\tilde x,t|x_0,0)} {\mathcal{D}}x(\tau)\, {\mathrm{e}}^{-\frac{1}{\epsilon}\int_0^t L(x(\tau),\dot x(\tau))\, d\tau}\,.
\end{equation}
Here, we denote by ${\mathcal{C}}(\tilde x,t|x_0,0)$ the set of all paths that are connecting the starting point $(x_0, 0)$ with the
end point $(\tilde x,t)$. The Lagrangian $L$ is given by 
\begin{equation}
L = \frac{1}{2} \langle \dot x - b(x), a^{-1}(\dot x - b(x))\rangle\,,
\end{equation}
where $\langle \cdot,\cdot \rangle$ denotes the usual inner product in ${\mathbb{R}}^n$ and the correlation matrix $a$ is
given by $a = \sigma\sigma^T$. For small $\epsilon$, intuition tells us that the transition probability should be dominated by
a path where the action $S$ defined as
\begin{equation}
S = \int_0^t L(x(\tau),\dot x(\tau))\, d\tau
\end{equation}
is minimal. This intuition can be made rigorous via using the Freidlin-Wentzell approach of large deviations \cite{freidlin-wentzell:1998}. In the mathematical
literature, a curve $\left(\tau,x(\tau)\right)$ that minimizes the action $S$ is called a {\em minimizer of the Freidlin-Wentzell action functional},
in the physics literature, such a path is often called an {\em instanton}. In a simple calculation, we can derive the associated Euler-Lagrange equations that an instanton needs to satisfy. However, in many instances, it is convenient to apply a Legendre-Fenchel transformation and
move from the Lagrangian to the Hamiltonian framework by introducing momenta $p_i = \partial L / \partial \dot x_i$.  For diffusive processes as
in (\ref{basic_sde}), the Hamiltonian $H$ is given by
\begin{equation}
H = \frac{1}{2} \langle p, ap\rangle + \langle b, p \rangle\,.
\end{equation}
Note that we restrict ourselves to processes where the drift $b$ and the noise $\sigma$ are not explicit functions of the time $t$. In this case, 
the Hamiltonian $H$ is conserved. One particular example that is of interest in many physical applications is the exit from a stable fixed point $x_s$ where the drift $b(x_s)$ is zero at the fixed point. Assume that we start from such a fixed point and we consider transitions to a different point $\tilde x$. It can be shown in general that the path that minimizes the action $S$ will need infinite time to move from the fixed point to the end point $\tilde x$ and
that this path corresponds to the constraint that $H=0$. Therefore, we are often interested in the Freidlin-Wentzell minimizer under the additional constraint that the Hamiltonian $H=0$. The minimizer or instanton satisfies the equations
\begin{equation} \label{instanton_general}
\dot x = \frac{\partial H}{\partial p} \equiv H_p, \qquad \dot p = - \frac{\partial H}{\partial x} = - H_x
\end{equation}
with the appropriate boundary conditions. It is common to parametrize the path of the instanton in a way that we set $x(\tau)$ to have the value $x_s$ for a time $\tau \to -\infty$ and to observe $x$ at time $\tau=0$ at $\tilde x$. Then, the boundary conditions for the instanton equations (\ref{instanton_general}) are given by
\begin{equation}
\lim_{\tau \to -\infty} x(\tau) = x_s, \qquad x(0) = \tilde x\,.
\end{equation}
There are only very few examples, where the instanton equations can be solved analytically. Especially for higher-dimensional systems, we need to find the minimizer of the functional $S$ numerically. Note that, in particular if the stochastic equation under consideration represents a finite-dimensional Galerkin approximation of a stochastic partial differential equation (SPDE), the dimension of $x$ can be large and efficient numerical methods are desirable. One numerically efficient approach is to parametrize the instanton using arc length instead of using the original parametrization in terms of the time $t$. This is possible if the drift $b$ and the covariance matrix $a$ do not explicitly depend on the time $t$ (but they can still depend on $x$). Using a more appropriate parametrization is the key idea of the {\em string method} \cite{e-ren-vanden-eijnden:2007} and the {\em geometric minimum action method} (gMAM) \cite{heymann-vanden-eijnden:2008,vanden-eijnden-heymann:2008}. 

Consider a parametrization $t = g(s)$ and let $\Phi(s) = x(g(s))$ be the path of the minimizer depending on the parameter $s$. Setting $\lambda(s) = 1/g'(s)$ we obtain using the chain rule and Hamilton's equations:
\begin{equation}
\lambda^2\Phi'' - \lambda H_{px}\Phi' + H_{pp}H_x + \lambda\lambda'\Phi' = 0\,.
\end{equation}
In order to find the minimizing path using gMAM, one introduces an artificial relaxation time $\tau$ to solve the equation
\begin{equation} \label{gMAMgeneral}
\dot \Phi \equiv  \frac{\partial\Phi}{\partial \tau} = \lambda^2\Phi'' - \lambda H_{px}\Phi' + H_{pp}H_x + \lambda\lambda'\Phi' + \mu \Phi'\,.
\end{equation}
The last term is used to enforce the constraint given by the parametrization with respect to arc length. 

\section{Schr{\"o}dinger systems and gMAM}

In this section we show how to implement the geometric minimum action method in the case of dissipative 
nonlinear Schr{\"o}dinger systems.

Consider the cubic nonlinear Schr{\"o}dinger equation for the complex field $A=A(z,t)$ given by
\begin{equation}
iA_z + dA_{tt} + c|A|^2A = -\alpha iA + i\kappa A_{tt} + \eta(z,t)\,.
\end{equation}
For a stochastic partial differential equation with a nonlinear drift operator ${\mathcal{B}}$ of the form (here $z$ is the evolution variable and $t$ is the transverse variable, ${\mathcal{W}}$ is a Brownian sheet)
\begin{equation}
\varphi_z = {\mathcal{B}}(\varphi)dz + \sqrt{\epsilon}d{\mathcal{W}}
\end{equation}
the equation (\ref{gMAMgeneral}) is written as \cite{heymann-vanden-eijnden:2008}
\begin{equation} \label{gMAM_nlse}
\dot \Phi = \lambda^2\Phi'' - \lambda\left(\partial{\mathcal{B}} - (\partial{\mathcal{B}})^+\right)\Phi' -  (\partial{\mathcal{B}})^+{\mathcal{B}} + \lambda'\lambda\Phi' + \mu \Phi'\,.
\end{equation}
In order to apply gMAM to the cubic nonlinear Schr{\"o}dinger equation, we need to compute the different terms in the equation above.
This can be done in the following way: Splitting the complex envelope $A=u(z,t)+iv(z,t)$ into real and imaginary part, we can write the nonlinear operator ${\mathcal{B}}$ as
\begin{equation}
{\mathcal{B}}(\Phi) =  \left(\begin{array}{c} {\mathcal{B}}_1(u,v) \\ {\mathcal{B}}_2(u,v) \end{array}\right) 
                              = \left(\begin{array}{c} b_1(u,v) \\ b_2(u,v) \end{array}\right) + \kappa  \left(\begin{array}{c} u_{tt} \\ v_{tt} \end{array}\right) 
                              + d  \left(\begin{array}{c} -v_{tt} \\ u_{tt} \end{array}\right)\,.
\end{equation}
For the NLSE in the above form, we find for $b_1$ and $b_2$ obviously
\begin{equation}
b_1(u,v) = -\alpha u - c(u^2+v^2)v, \qquad b_2(u,v) = -\alpha v - c(u^2+v^2)u \,.
\end{equation}
The linearization $\partial{\mathcal{B}}$ of the nonlinear operator ${\mathcal{B}}$ is given by
\begin{equation} \label{split_B}
\partial{\mathcal{B}} = \nabla b + \kappa {\mathcal{L}}_D + d{\mathcal{L}}_H
\end{equation}
where the operators $ {\mathcal{L}}_D $ and ${\mathcal{L}}_H$ are defined by
\begin{equation}
{\mathcal{L}}_D = \left(\begin{array}{cc} \partial_t^2 & 0 \\ 0 & \partial_t^2 \end{array}\right), \qquad 
{\mathcal{L}}_H = \left(\begin{array}{cc} 0 & -\partial_t^2 \\  \partial_t^2 & 0 \end{array}\right)\,.
\end{equation}
With these definitions, we can write for the $\partial{\mathcal{B}} - (\partial{\mathcal{B}})^+$ 
\begin{equation}
\partial{\mathcal{B}} - (\partial{\mathcal{B}})^+ = \nabla b - (\nabla b)^T + 2d {\mathcal{L}}_H
\end{equation}
and for the term $(\partial{\mathcal{B}})^+{\mathcal{B}}$ we find
\begin{equation}
(\partial{\mathcal{B}})^+{\mathcal{B}} = (\nabla b)^T{\mathcal{B}} + \left(\kappa {\mathcal{L}}_D - d {\mathcal{L}}_H\right) b + (\kappa^2+d^2)\partial_t^4 \Phi\,.
\end{equation}
In the numerical implementation, it is essential to treat carefully the last term involving fourth-order derivatives $(\kappa^2+d^2)\partial_t^4 \Phi$ in order 
to avoid instabilities. This is the reason, why we split up the second-order derivative terms in the representation (\ref{split_B}) of ${\mathcal{B}}$. Summarizing
the results, for the components $\Phi = (u,v)^T$, we obtain the two equations
\begin{eqnarray}
\dot u &=& \lambda^2u'' - \lambda\left(b_{1v}-b_{2u} -2dv_{tt}'\right) \nonumber \\ && - \left(b_{1u}\mathcal{B}_1 + b_{2u} \mathcal{B}_2 + \kappa b_{1tt} + db_{2tt} + (\kappa^2+d^2)u_{tttt}\right) \,, \\
\dot v &=& \lambda^2v'' - \lambda\left(b_{2u}-b_{1v} -2du_{tt}'\right) \nonumber \\ && - \left(b_{1u} \mathcal{B}_1 + b_{2v} \mathcal{B}_2 + \kappa b_{2tt}   - db_{1tt} + (\kappa^2+d^2)v_{tttt}\right) \,.
\end{eqnarray}
%
%
%
\begin{figure}[htb!]
\centering
\hfill
\includegraphics[]{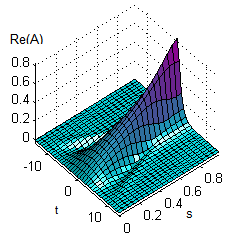}
\hfill
\includegraphics[]{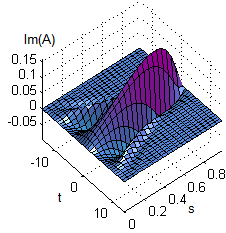}
\hfill
\caption{Evolution of the instanton of the linear dissipative Schr{\"o}dinger equation with respect to arc length parametrization. The real part of the instanton is significantly larger than its imaginary part, which has to be zero at both boundaries.} 
\label{fig:LingMAM}
\end{figure} 

\section{Fourier domain solution of the linear case}

In order to test the numerical implementation of gMAM for Schr{\"o}dinger systems, we can consider the particular case of a linear system. The instanton equations
for the minimizer of the Freidlin-Wentzell action functional are, in this case, given by
\begin{equation} \label{lin_schroedinger}
A_z = -\alpha A + \kappa A_{tt} + idA_{tt} + P\,, \qquad P_z =  \alpha P - \kappa P_{tt}  + idP_{tt}\,.
\end{equation}
Note that the equation of the optimal noise field $P$ does not contain any term involving the field $A$. In a typical gMAM setting, however, we consider
transitions from a state $A_1$ of the field $A$ to another state $A_2$, leading to boundary conditions for the field $A$. The minimizer has to satisfy these
boundary conditions which will impose boundary conditions on the field $P$. These boundary conditions are only known {\em after} the computation of the
minimizer $A$. The situation is different if we prescribe an initial condition for the field $A$ and a final condition for the field $P$ \cite{chernykh-stepanov:2001,grafke-grauer-schaefer-etal:2014}. 
Then, in the linear case above, the equation for $P$ can be solved 
independently of the equation for $A$ and the solution can be used to solve the equation for $A$. As a test case, let us choose the following boundary conditions
for $P$:
\begin{equation}
\lim_{z \to -\infty} P(z,t) = 0, \qquad P(0,t) = F\delta(t)\,.
\end{equation}
Note that it can be shown from the variational principle that this final condition for the noise field $P$ corresponds to a final condition $A(0,t) = a$ for the field
$A$ with $F$ acting as Lagrange multiplier to enforce this constraint.
Using Fourier transform, we can immediately solve (\ref{lin_schroedinger}) and obtain the solution for the Fourier transform $\hat A$ of the field $A$:
\begin{equation}
\hat A(z,\omega) = \frac{F}{2(\alpha + \kappa \omega^2)}\,{\mathrm{e}}^{(\alpha+\kappa\omega^2 - id\omega^2)z}
\end{equation}
At $z=0$, it is simple to carry out the inverse transform analytically and to obtain an explicit relationship between the amplitude $a$ and the Lagrange multiplier $F$:
\begin{equation}
A(0,t) = \frac{F}{4\sqrt{\alpha\kappa}}\,{\mathrm{e}}^{-\sqrt{\alpha/\kappa}|t|}, \qquad F = 4\sqrt{\alpha\kappa}\,a\,.
\end{equation}
In this way, we can use this linear case as a test case for gMAM since both the initial state $\lim_{z\to -\infty} A(z,t)=0$ and the final state $A(0,t)$ are known.

\begin{figure} [htb]
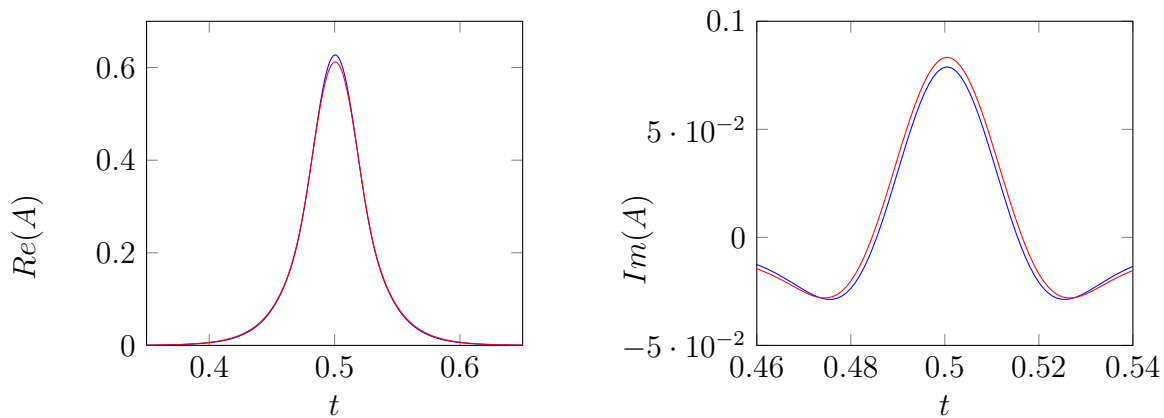

\centering 
\setlength\figureheight{4cm} 
\setlength\figurewidth{5cm} 
\input{UvUanaLin0.tikz} \hfill \input{VvVanaLinTest0.tikz} 
\caption{Comparison between a slice of the the analytic instanton calculation for the linear Schr{\"o}dinger equation (red) against the gMAM result (blue) for arc length parameter $s = .78$. The left panel shows the real part $u$, the right panel the imaginary part $v$. The small error decreases with refinement of the computational grid.}
\label{fig:LinComp} 
\end{figure}

Fig$.$ \ref{fig:LingMAM} shows the evolution of the instanton (real and imaginary part) with respect to arc length. Note that, if we set the dispersion $d=0$, we would
obtain a solution with an imaginary part equal to zero. Also, for the dispersive case, we observe in both real and imaginary part, slow oscillations with respect to the transversal coordinate $t$ 
at the beginning of the instanton's evolution. In order to check the accuracy of the numerical code, we can compare the solution obtained by numerically solving (\ref{gMAM_nlse}) to 
the analytical solution obtained in Fourier space. Fig$.$ \ref{fig:LinComp} shows the comparison of a slice at a fixed arc length $s$.

\section{Minimizer of the cubic nonlinear Sch{\"o}dinger equation}

In the following we are looking at the nonlinear case, where we set for simplicity the dispersion coefficient $d=1/2$ and the nonlinear coefficient $c=1$.
It is well-known that, for the lossless case without stochastic perturbations, the cubic nonlinear Sch{\"o}dinger equation is integrable and possesses soliton
solutions \cite{newell-moloney:1992,hasegawa-kodama:1995}. As a simple example, we look at the stochastic transition of a 'flat soliton' to a 'sharply peaked' soliton, hence we choose
as boundary conditions $A_k (t) = \Lambda_k/\cosh(\Lambda_k t)$ with $\Lambda_1 < 1 < \Lambda_2$. While we do not have an analytical solution to the
instanton equations in this case, we can still use them to verify the accuracy of the gMAM algorithm. The minimizer $(A,P)$ has to satisfy the pair of coupled Euler-Lagrange equations
\begin{eqnarray} \label{instanton_eqs_nlse}
A_z &=& - \alpha A + \kappa A_{tt} + \frac{i}{2} A_{tt} + i|A|^2A + P\,, \\
P_z &=&   \alpha P - \kappa P_{tt}  + \frac{i}{2} P_{tt} + 2i|A|^2P - iA^2P^*\,.
\end{eqnarray}
From the numerical solutions given by gMAM, we can take the initial condition $A_1$ and propagate this initial condition using the evolution equation of $A_z$ (forward in $z$-direction).
In a similar fashion we can take the final condition $P_2$ and propagate this final condition using the evolution equation of $P_z$ (backward in $z$-direction). Note that, when solving these equations,
it is appropriate to scale $z$ as well according to arc length - and the corresponding transformation is also provided by gMAM algorithm in terms of the parameter $\lambda(s)$. The following
fig. \ref{fig:solcomp} shows contour plots for the instanton of the transition of a soliton with $\Lambda_1 = 0.5$ to $\Lambda_2=2.5$. The dissipative coefficients $\alpha$ and $\kappa$ are 
both set to 0.1 in this simulation. The left panel shows the gMAM results and the right panel the solutions
of the system of equations (\ref{instanton_eqs_nlse}) with the initial condition $A_1$.

\begin{figure}[htb!]
\centering
\setlength\figureheight{5cm} 
\setlength\figurewidth{6cm} 
%
%
\begin{tikzpicture}

\begin{axis}[%
width=\figurewidth,
height=\figureheight,
axis on top,
scale only axis,
xmin=-25.09765625,
xmax=24.90234375,
xlabel={time t},
y dir=reverse,
ymin=-0.00196078431372549,
ymax=1.00196078431373,
ylabel={arc length s}
]
\addplot [forget plot] graphics [xmin=-25.09765625,xmax=24.90234375,ymin=-0.00196078431372549,ymax=1.00196078431373] {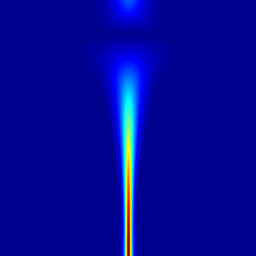};
\end{axis}
\end{tikzpicture}%
\hfill
%
%
\begin{tikzpicture}

\begin{axis}[%
width=\figurewidth,
height=\figureheight,
axis on top,
scale only axis,
xmin=-25.09765625,
xmax=24.90234375,
xlabel={time t},
y dir=reverse,
ymin=-0.00196078431372549,
ymax=1.00196078431373,
ylabel={arc length s}
]
\addplot [forget plot] graphics [xmin=-25.09765625,xmax=24.90234375,ymin=-0.00196078431372549,ymax=1.00196078431373] {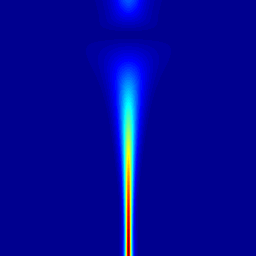};
\end{axis}
\end{tikzpicture}%
\caption{Contour plot of the amplitude $|A|$ of the minimizer of the nonlinear Schr{\"o}dinger equation: Left panel shows the result of the gMAM computation. Right panel shows the solution of the instanton equation
of $A$.}  
\label{fig:solcomp}
\end{figure} 

\section{Comparison to a finite-dimensional model}

\begin{figure}[htb!]
\centering
\setlength\figureheight{5cm} 
\setlength\figurewidth{6cm} 
%
%
\begin{tikzpicture}

\begin{axis}[%
width=\figurewidth,
height=\figureheight,
axis on top,
scale only axis,
xmin=-25.09765625,
xmax=24.90234375,
xlabel={time t},
y dir=reverse,
ymin=-0.00196078431372549,
ymax=1.00196078431373,
ylabel={arc length s}
]
\addplot [forget plot] graphics [xmin=-25.09765625,xmax=24.90234375,ymin=-0.00196078431372549,ymax=1.00196078431373] {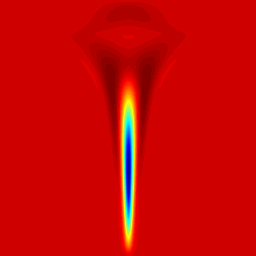};
\end{axis}
\end{tikzpicture}%
\hfill
%
%
\begin{tikzpicture}

\begin{axis}[%
width=\figurewidth,
height=\figureheight,
axis on top,
scale only axis,
xmin=-25.09765625,
xmax=24.90234375,
xlabel={time t},
y dir=reverse,
ymin=-0.00196078431372549,
ymax=1.00196078431373,
ylabel={arc length s}
]
\addplot [forget plot] graphics [xmin=-25.09765625,xmax=24.90234375,ymin=-0.00196078431372549,ymax=1.00196078431373] {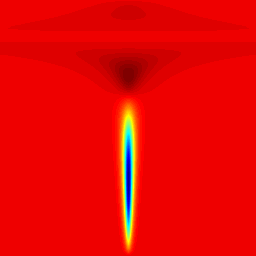};
\end{axis}
\end{tikzpicture}%
\caption{Contour plot of the imaginary part $v$ of the minimizer of the nonlinear Schr{\"o}dinger equation (left panel) and the approximation using the finite-dimensional set
of stochastic equations (right panel).}
\label{fig:mincomp}
\end{figure}

\begin{figure}[htb!]
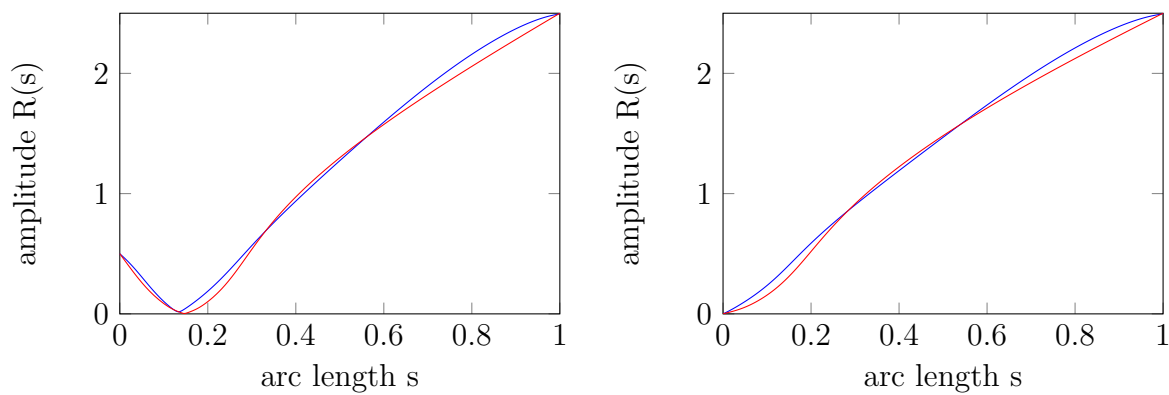

\centering
\setlength\figureheight{4cm} 
\setlength\figurewidth{5.85cm} 
\input{flat2peak.tikz} 
\hfill
\input{zero2peak.tikz} 
\caption{Left panel: plot of the amplitude $R(s)$ of the ODE minimizer (red) in comparison to the amplitude of the PDE minimizer (blue) for the transition of a soliton with $\Lambda_1 = 0.5$ to 
$\Lambda_2 = 2.5$. Right panel:  same comparison but for the exit from zero the the soliton with an amplitude $\Lambda_2 = 2.5$.}
\label{fig:plot_a}
\end{figure}

Over the last decades, in particular in the context of the cubic nonlinear Schr{\"o}dinger equation, much work has been devoted on the study of finite-dimensional models as 
approximations to infinite dimensional models.
In such a setting, the original stochastic partial differential equation is approximated by a low-dimensional random dynamical system \cite{abdullaev-bronski-etal:2000,schaefer-moore-etal:2002}. In many instances, however, the question whether results obtained
from the analysis of the reduced model are still valid for the full system given by the partial differential equation has not been studied thoroughly \cite{Moore_Schafer_Jones_2005}. In the following, we work with 
a finite-dimensional reduction from a recent paper by R. Moore \cite{moore:2014}, adapted for our case: We parametrize the soliton's evolution by four parameters $(R,\Omega,T,\phi)$ that are
all functions of the evolution variable $z$. The corresponding pulse $A$ can be constructed from these parameters via
\begin{equation} \label{A_param}
A(z,t) = \frac{R(z)}{\cosh\left(R(z)(t-T(z)\right)}\,{\mathrm{e}}^{i\phi(z)+it\Omega(z)}\,.
\end{equation}
The stochastic dynamical system of the parameters is given by 
\begin{equation}
\left(\begin{array}{c} dR \\ d\Omega \\ dT \\ d\phi \end{array}\right) = 
\left(\begin{array}{c} -2\alpha R - \frac{2}{3}\kappa R^3 - 2\kappa R\Omega^2 \\ 
                                -\frac{4}{3}\kappa R^2\Omega \\ \Omega \\
                                 \frac{1}{2} R^2 - \frac{1}{2} \Omega^2 + \frac{4}{3}\kappa R^2 \Omega T \end{array}\right) dz 
+ \sqrt{\epsilon} \Sigma \left(\begin{array}{c} dW_1 \\ dW_2 \\ dW_3 \\ dW_4 \end{array} \right)
\end{equation}
together with the correlation matrix $\Sigma\Sigma^T$ which reads
\begin{equation}
\Sigma\Sigma^T = 2\left(\begin{array}{cccc} R & 0 &-T & 0 \\ 0 & \frac{R}{3} & 0 &-\frac{RT}{3} \\ -T & 0 & \frac{2T^2}{R} + \frac{\pi^2}{12R^3} & 0 \\ 
0 & -\frac{RT}{3} & 0 & \frac{RT^2}3 + \frac{\pi^2+12}{36R} 
\end{array}\right) \,.
\end{equation}
In this four-dimensional reduction, the soliton transition discussed earlier corresponds to a transition of the states $(\Lambda_1,0,0,0)$ to $(\Lambda_2,0,0,0)$,
hence for the boundary conditions we can choose $R_1=\Lambda_1$ and $R_2=\Lambda_2$. In order to compare the results of this approach to the PDE minimizer,
we first use gMAM to compute the minimizing path $(R(s),\Omega(s),T(s),\phi(s))$ and then we construct the approximation of the PDE minimizer via (\ref{A_param}).
Note that it is convenient to parametrize the approximate PDE minimizer with respect to arc length such that it can be compared to the minimizer from the PDE
model. Fig$.$ \ref{fig:mincomp} shows as an example the comparison of the imaginary part of the PDE minimizer to the approximation obtained by the stochastic ODEs. 
At the beginning of the evolution, there are again slow oscillations (similar to the linear case) which cannot be properly captured by the finite-dimensional
model. For larger $s$, the agreement of the two solutions is remarkable.
While a detailed comparison between minimizers of the finite-dimensional reduction and the full model is beyond the scope of the present work, we remark that for 
the example studied here (in particular for the chosen parameters), the major contribution stems from the evolution equation of the amplitude $R(z)$. As a first step,
we can simply extract the maximum value of the pulse profile $|A|$ for each value of the arc length $s$. The corresponding graph is shown on the left panel of fig. \ref{fig:plot_a}. 
Again, we observe a good agreement between the ODE prediction and the PDE minimizer. Note that, initially the amplitude dips to a fairly low value 
(this can already be seen in the contour plots shown previously in fig. \ref{fig:solcomp}). The ODE model captures this dip fairly well. 

\begin{figure} [htb!]
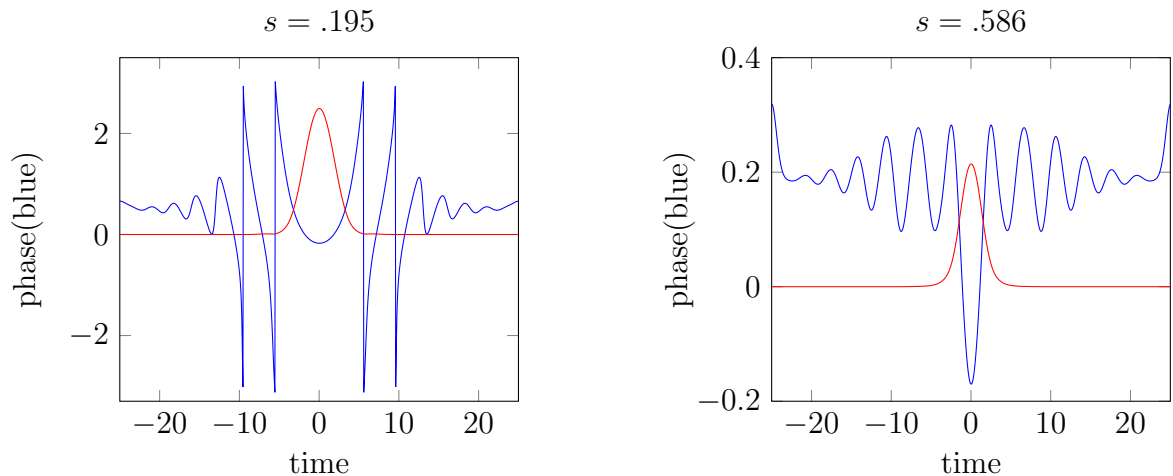

\centering 
\setlength\figureheight{4cm} 
\setlength\figurewidth{5.3cm} 
\input{pPhaseWithAmp2.tikz} \hfill \input{pPhase3WithAmp05.tikz} 
\caption{Example plots of phase vs time (in blue) plotted on the same axes with a magnified plot of the amplitude vs time (in red) for s=.195 (left) and s = .586 (right). We can see that the phase shows parabolic behavior within the time frame where the amplitude is large which could justify a using a parabolic phase model to characterize this system.}  \label{fig:pPhase} 
\end{figure}

A further important example is the transition from the zero state to a soliton. Note that this transition cannot be captured precisely in the ODE model, as the equations become singular in the limit $R\to 0$. 
However, we can compute the ODE minimizer starting from a fairly small amplitude $\Lambda_1 = \delta \ll 1$. The result is presented on the right panel of fig. \ref{fig:plot_a}. In this example,
we chose $\delta =  0.001$. Again, the amplitude of the ODE minimizer and the amplitude of the PDE minimizer show good agreement. While these agreements are encouraging and clearly support validity of the finite-dimensional model to capture the transition of solitary waves of different amplitudes, we also noticed differences between the PDE minimizer and the ODE minimizer: When looking at the pulse shape of the PDE minimizer and as seen in fig$.$ \ref{fig:pPhase}, we find a presence of a parabolic phase  (often called "chirp" in the optics) which is not captured by the ODE model above.   A thorough analysis of the potential impact of the chirp, including the possibility to extend the low-dimensional model to include a parabolic phase, is beyond the present paper and subject to future research.

\section{Conclusion}

This paper shows how to adapt and implement the geometric minimum action method (gMAM) for the case of a stochastic cubic nonlinear Schr{\"o}dinger equation.
The resulting implementation was tested using an analytical solution for the linear case and an independent solution based on direct integration of the instanton equation for the nonlinear case. We applied this method to the computation of the minimizer for the transition of solitons with small or zero amplitude to peaked solitons with larger amplitude. Finally, the PDE minimizer was compared to the minimizer of a low-dimensional approximation.

\section*{Acknowledgments}
The authors thank R. Moore, E. Vanden-Eijnden, T. Grafke, R. Grauer,
and S. Turitsyn for many valuable discussions. 
This work was supported, in part, by the NSF grants DMS-1108780
and DMS-1522737.

\section*{References}

\bibliography{master}

\end{document}